\documentclass[twocolumn,showpacs,preprintnumbers,amsmath,amssymb,floatfix]{revtex4}
\usepackage{subfigure}
\usepackage{amsmath}
\usepackage{amsfonts}
\usepackage{amssymb}
\usepackage[dvips]{graphicx}

\usepackage[colorlinks]{hyperref}
\usepackage[usenames,dvipsnames]{color}

\usepackage{bbm}
\def\beq{\begin{equation}}
\def\eeq{\end{equation}}
\def\bea{\begin{eqnarray}}
\def\eea{\end{eqnarray}}
\def\ba{\begin{array}}
\def\ea{\end{array}}

\def\part{\partial}

\begin{document}

\preprint{UdeM-GPP-TH-10-189}

\title{Fate of the false monopoles: induced vacuum decay}
\author{Brijesh Kumar$^{1,2}$}
\email{brijesh@phy.iitb.ac.in}
\author{M. B. Paranjape$^2$}
\email{paranj@lps.umontreal.ca}
\author{U. A. Yajnik$^{1,2,3}$}
\email{yajnik@iitb.ac.in}

\affiliation{$^1$ Physics Department, Indian Institute of Technology Bombay, Mumbai, 400076, India}
\affiliation{$^2$Groupe de physique des particules, D\'epartement de
physique, Universit\'e de Montr\'eal, C.P. 6128, Succ. Centre-ville,
Montr\'eal, Qu\'ebec, CANADA, H3C 3J7 }
\affiliation{$^3$ Department of Physics, Ernest Rutherford Physics Building, McGill University, 3600 rue University, 
Montr\'eal, Qu\'ebec, CANADA, H3A 2T5}

\begin{abstract}
We study a gauge theory model where there is an intermediate symmetry breaking to a meta-stable vacuum that breaks a simple gauge group to a $U(1)$ factor.  Such models admit the existence of meta-stable magnetic monopoles, which we dub false monopoles.   We prove the existence of these monopoles in the  thin wall approximation.  We determine the instantons for the collective coordinate that corresponds to the radius of the monopole wall and we calculate the semi-classical tunneling rate for the decay of these monopoles.  The monopole decay consequently triggers the decay of the false vacuum.   As the monopole mass is increased, we find an enhanced rate of decay of the false vacuum relative to the celebrated homogeneous tunneling rate due to Coleman \cite{Coleman:1975qj}.
\end{abstract}

\maketitle
\section{Introduction}

Semi-classical solutions with topologically non-trivial boundary conditions in relativistic 
field theory \cite{'tHooft:1974qc,Polyakov:1974ek,Polyakov:1976fu} have the interesting property that
they interpolate between two or more alternative translationally invariant vacua of the theory. 
For instance the exterior of a monopole or a vortex solution is a phase of broken
symmetry while the interior of the object generically contains a limited region of
unbroken symmetry (for more details and lucid expositions see \cite{Coleman:1975qj} and \cite{Rajaraman:1982is}). 
Most of the commonly studied solutions are topologically non-trivial, however non-trivial boundary
conditions are not a guarantee of dynamical stability. In \cite{Preskill:1992ck} for example a large
number of such solutions are constructed in gauge field theories which are generically metastable. The skyrmion is also a classic example of a topologically non-trivial configuration that is unstable without the addition of fourth order Skyrme term \cite{Skyrme:1962vh,Gisiger:1998tv}.  
All of the classically stable solutions (allowing for quantum metastabilty), are non-trivial time independent local minima of the effective action of the theory.

The metastability of such solutions can be of significant interest. The implied decay of the
object would be accompanied by the change in phase of the system as a whole. In the context of
cosmology this may imply a change  in the cosmic history and determine the abundance of
relic objects.  On a more formal footing the question of metastability of vacua has
gained considerable interest in the context of supersymmetric field theories \footnote{See \cite{Terning:2003th} 
for discussion of supersymmetric field theories.} where
a non-supersymmetric phase is required on phenomenological grounds but such a phase
is necessarily metastable on theoretical grounds \cite{Dine:1993yw,Intriligator:2006dd}. 
In String Cosmology the de Sitter solution obtained is generically meta-stable \cite{Kachru:2003aw} 
and its phenomenological viability depends on the tunneling rate being sufficiently slow.

Change in  phase due to metastable topological objects is a generalisation of the 
following better known mechanism.
When the effective potential of the theory possesses several local minima, all but the
lowest minimum are quantum mechanically unstable. The so called false vacua are then
liable to decay, even in the absence of topological objects, according to a rate given by 
a WKB-like formula studied earlier by 
\cite{Kobzarev:1974cp} and provided an elegant and lucid footing by Coleman \cite{Coleman:1977py,Coleman:1980aw}.
The cases studied there concerned a transition between two translationally invariant vacua. 
The generic scenario of decay consists of spontaneous formation of a small \textit{bubble} of true vacuum, 
which can then start growing by semi-classical evolution. In Minkowski space, the formation of 
one such bubble  is sufficient to convert the phase of the system to the true vacuum.
In the context of an expanding Universe, conversion of the entire Universe to the true vacuum
would require formation of sufficiently large number of such bubbles at an adequate rate.

The existence of topological objects may provide additional sources of metastability.
Phase transitions seeded by topological solutions were studied early in the works of
\cite{Steinhardt:1981ec,Hosotani:1982ii,Yajnik:1986tg,Yajnik:1986wq}.  An essential aspect of
these studies is precisely the observations that there exist solutions with non-trivial
boundary conditions which interpolate between two distinct minima of the effective potential.
The importance of this alternative route to decay arises from the fact that it can be
much more rapid than the spontaneous decay of a translationally invariant vacuum. Indeed, for some values of the parameters the decay induced by topological objects may require no tunneling
and therefore would be very prompt in a context where the parameters are changing adiabatically, 
as for instance in the early Universe.

Obtaining a general formula characterising this kind of vacuum decay has been rather elusive
although the ideas have been adequately explicated in \cite{Steinhardt:1981ec,Hosotani:1982ii,
Yajnik:1986tg,Yajnik:1986wq}.
More recently, the relevance of the mechanism has been demonstrated in specific examples, 
in \cite{Kumar:2008jb} for the mediating sector of a hidden sector scenario of supersymmetry 
breaking and in \cite{Kumar:2009pr} in a GUT model with O'Raifeartaigh type direct supersymmetry 
breaking. In this paper we explore a model that is amenable to an analytical treatment within 
the  techniques developed in \cite{Coleman:1978ae}. In doing so we provide a transparent model 
in which the generic expectations raised in \cite{Steinhardt:1981ec,Hosotani:1982ii,
Yajnik:1986tg,Yajnik:1986wq} can be realised 
and a specific formula can be derived.

We construct an $SU(2)$ gauge model with a triplet scalar field with two possible translationally 
invariant vacua, one with $SU(2)$ broken to $U(1)$ and the other with the original gauge symmetry 
intact. The former phase permits the existence of monopoles. By appropriate choice of potential
for the triplet it can be arranged that the phase of unbroken symmetry is lower in energy and
represents the true vacuum of the theory. The monopoles interpolate between the true vacuum and
the false vacuum.  For a wide range of the parameters, these monopoles are in fact classically stable.  In previous work \cite{Steinhardt:1981ec,Hosotani:1982ii}  the dissociation of such monopoles was considered, varying the parameters of the theory to critical values where the monopoles were classically unstable due to infinite dilation.  This can occur for example in the early Universe where the high temperature phase prefers one
vacuum in which the system starts, but with adiabatic reduction in temperature, a different 
phase becomes more favorable. The Universe is then
liable to simply \textit{roll over}, by classical evolution, to the true vacuum.

It was however, overlooked that  these monopoles are in fact unstable due to quantum tunneling well before the parameters reach their critical values.  We dub such monopoles \textit{false monopoles}.  
Working in the thin wall limit for the monopoles \cite{Steinhardt:1981ec}, we show that such monopoles undergo quantum tunneling to larger monopoles, which are then classically unstable by expanding indefinitely, consequently converting all space to the true vacuum, the phase of unbroken $SU(2)$ symmetry. Further, the
formula we derive also recovers the regime of parameter space, within the thin wall monopole limit,  where no tunneling is required for the decay but the monopole is simply classically unstable as previously treated \cite{Steinhardt:1981ec,Hosotani:1982ii}.

The rest of the paper is organised as follows. In section \ref{sec:model} we specify the model under
consideration and the monopole ansatz along with the equations of motion.  In  section \ref{sec:thinwallmonopole} we delineate the conditions under in which there should exist a metastable
monopole solution with a large radius and a thin wall.  We find the thin wall monopole solutions and also justify their existence.  In section \ref{sec:collco}  we use the thin wall approximation which permits a treatment of the solution in terms of a single collective coordinate, the
radius $R$ of the thin wall.   We argue that the monopole is unstable to
tunneling to a new configuration of a much larger radius and we determine the existence of the instanton
for this tunneling within the same thin wall approximation. In section \ref{sec:decay} we determine
the Euclidean action for this instanton, the so called bounce $B$ which determines
the tunneling rate for the appearance of the large radius unstable monopole.  In section \ref{sec:gutmodel} we relate our findings to a previous study of classical monopole instability in supersymmetric GUT models.   In section \ref{sec:conc} we discuss our results and compare our tunneling rate formula with that of the 
homogeneous bubble formation case  without monopoles. We show that in addition to our tunneling rate being significantly
faster, it also indicates a regime in which the monopoles become unstable,
hence showing that the putative non-trivial vacuum indicated by the effective potential is in fact unstable.

\section{Unstable monopoles in a false vacuum}
\label{sec:model}

\begin{figure}[!htp]
\begin{center}
\includegraphics[width=0.45\textwidth]{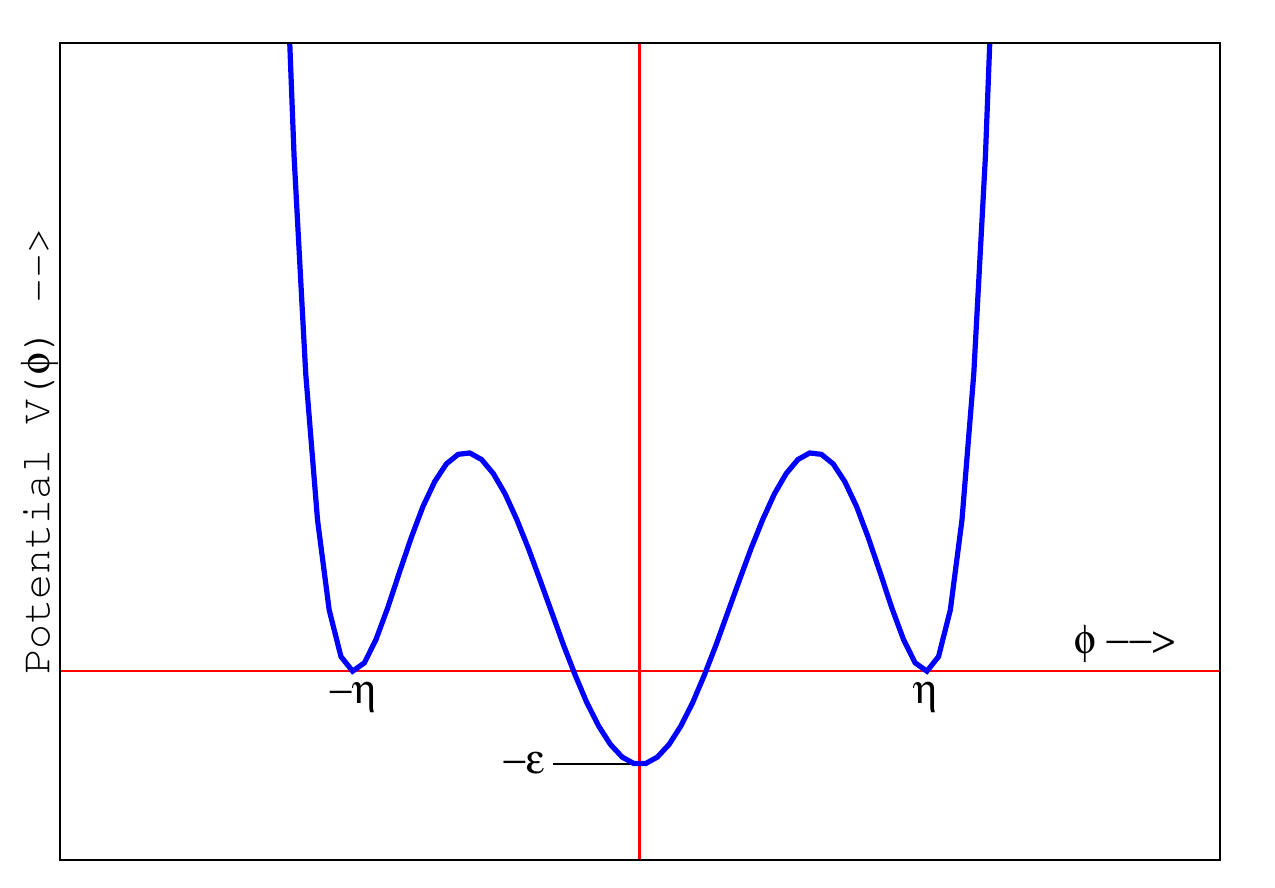}
\caption{The potential $V(\phi)$ for $\gamma \neq 0$ as a function of one of the components of the field $\phi$,
shifted by an additive constant so that $\phi = \eta$ has vanishing $V$ and the true
vacuum has $V = -\epsilon$.}
\label{fig1}
\end{center}
\end{figure}

Consider an $SU(2)$ gauge theory with a triplet scalar field $\phi$ with the Lagrangian density given by
\begin{equation}
\mathcal{L} = -\frac{1}{4}F_{\mu\nu}^a F^{\mu\nu a} + \frac{1}{2}(D_\mu \phi^a)(D^\mu \phi^a) - V(\phi^a\phi^a)
\end{equation} 
where 
\begin{equation}
F_{\mu\nu}^a = \partial_\mu A_\nu^a - \partial_\nu A_\mu^a + e\epsilon^{abc}A_\mu^b A_\nu^c,
\end{equation} 
and
\begin{equation}
D_\mu \phi^a = \partial_\mu \phi^a + e\epsilon^{abc} A_\mu^b \phi^c.
\end{equation}
The potential we use is a polynomial of order $6$ in $\phi$ and may conveniently be written as
\begin{equation}
V(\phi) = \lambda \phi^2 (\phi^2 - a^2)^2 + \gamma^2 \phi^2 -\epsilon
\label{potential}
\end{equation} 
where $\epsilon$ is defined so that the potential vanishes at the meta-stable vacua.    The vacuum energy density difference is then equal to $\epsilon$.
Such a potential was numerically analyzed by \cite{Steinhardt:1981mm}  as a toy model for the dissociation of monopoles.  Here we obtain explicit analytical formulae for the quantum tunneling decay of the monopoles.  
The potential has a minimum  at $\phi^T\phi=0$ which for $\gamma=0$ is degenerate with the manifold
of vacua at  $\phi^T\phi = a^2$. When we set $\gamma \neq 0$, we get a manifold of degenerate metastable 
vacua  at $\phi^T\phi = \eta^2$ (where the exact value of the VEV, $\eta$, is calculable and satisfies $\eta\approx a$ for small $\gamma$), and the  minimum at 
$\phi = 0$ becomes the true vacuum.
A plot of the potential for small $\gamma$ as a function of one of the components of $\phi$ is 
shown in figure \ref{fig1}.    A supersymmetry breaking model \cite{Bajc:2008vk} containing monopoles and 
a scalar potential similar to the one given in Eqn. (\ref{potential}) was studied in \cite{Kumar:2009pr}.

The manifold of vacua at $\phi^T\phi=\eta^2$ is topologically an $S^2$ and as spatial infinity is topologically also $S^2$, the appropriate homotopy group of the manifold of the vacua of the
symmetry breaking $SU(2)\rightarrow U(1)$ is $\Pi_2(SU(2)/U(1))$ which is $\mathbf{Z}$. This suggests the existence 
of topologically non-trivial solutions of the monopole type which are classically stable. The presence 
of the  global minimum at $\phi=0$ allows for the possibility that the monopole solution although topologically non-trivial, could be dynamically unstable.

A time independent spherically symmetric ansatz for the monopole can be chosen in the usual way as 
\begin{eqnarray}
\phi_a &=& \hat{r}_a \, h(r) \nonumber \\
A_\mu^a &=& \epsilon_{\mu ab}\,\hat{r}_b \,\frac{1 - K(r)}{er} \nonumber \\
A_0 &=& 0
\end{eqnarray} 
where $\hat{r}$ is a unit vector in spherical polar coordinates 
The energy of the 
monopole configuration in terms of the functions $h$ and $K$ is
\begin{align}
E(K,h) = 4\pi \int_0^\infty dr \Big( &\frac{(K')^2}{e^2} + \frac{(1-K^2)^2}{2e^2r^2} + \frac{1}{2}r^2(h')^2 \nonumber \\
&+ K^2h^2 + r^2V(h) \Big)
\label{staticenergy}
\end{align} 
where derivatives with respect to $r$ are denoted by primes. The static monopole solution is the minimum 
of this functional and the ansatz functions satisfy the equations
\begin{eqnarray}
h'' + \frac{2}{r} h' - \frac{2h}{r^2}K^2 - \frac{\partial V}{\partial h} &=& 0 \label{monopole-equation} \\
K'' - \frac{K}{r^2}(K^2 - 1) - e^2h^2K &=& 0.
\label{eom}
\end{eqnarray} 
As $r\rightarrow \infty$ the function $h$ asymptotically approaches $\eta$ and is 
zero at $r = 0$ from continuity requirements. On the other hand, $K$ approaches zero at spatial infinity 
so that the gauge field  decreases as $1/r$, and $K = 1$ at $r = 0$.

\begin{figure}[!htp]
\begin{center}
\includegraphics[width=0.45\textwidth]{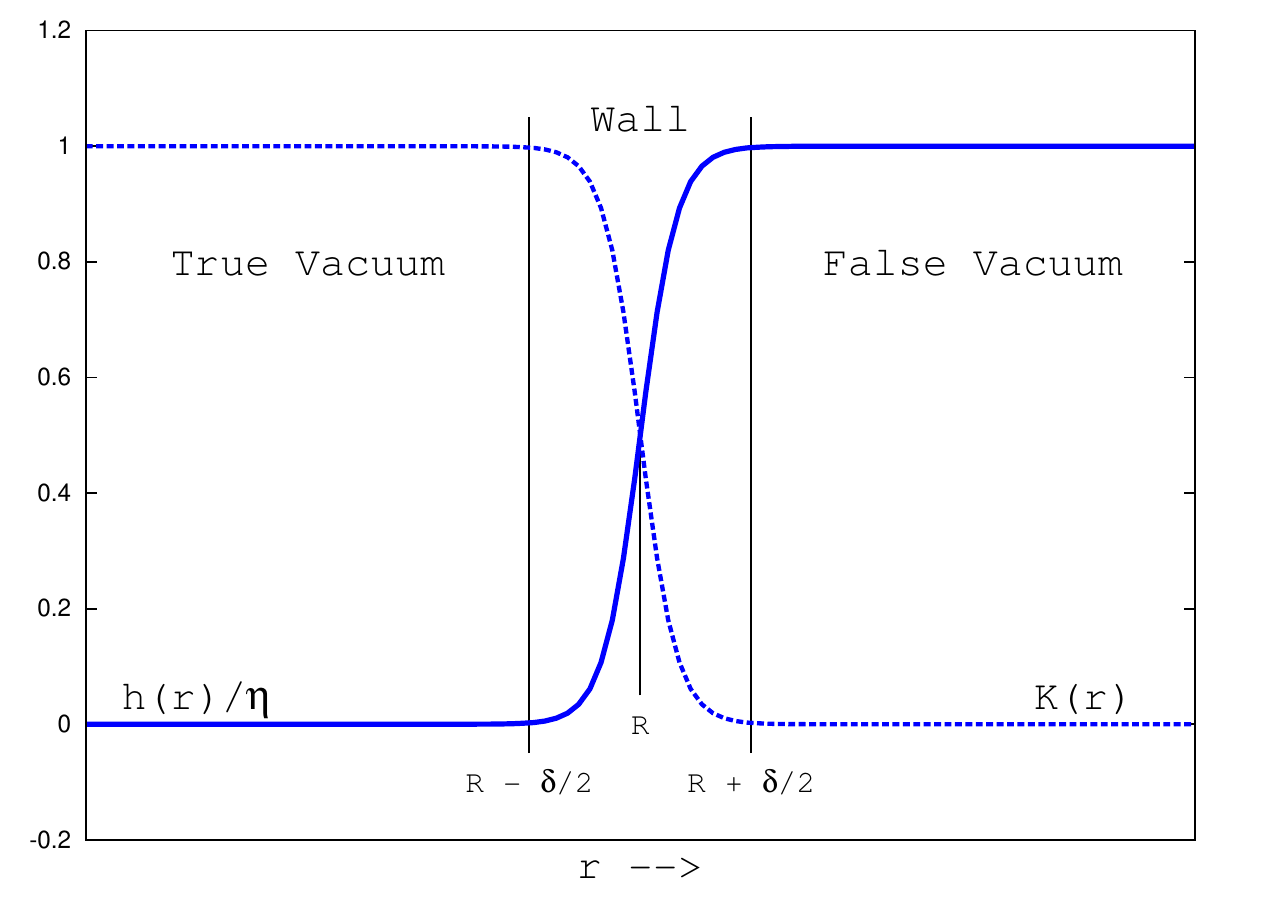}
\caption{The monopole profile under the thin wall approximation.}
\label{fig2}
\end{center}
\end{figure}

\section{Thin walled monopoles}
\label{sec:thinwallmonopole}

When the difference between the false and true vacuum energy densities $\epsilon$ is small, the monopole can be
treated as a thin shell, the so called thin wall approximation. Within this approximation, the monopole 
can be divided into three regions as shown in figure \ref{fig2}.
There is a region of essentially true vacuum extending from $r = 0$ upto a radius $R$. At $r = R$, there is a thin shell of thickness 
$\delta$ in which the field value changes exponentially from the true vacuum to the false vacuum. Outside this shell the monopole is essentially in the false vacuum, and so we have
\begin{eqnarray}
h \approx 0 \,,\,  K \approx 1 \qquad \qquad r < R - \frac{\delta}{2} \nonumber \\
h \approx \eta \,,\, K \approx 0 \qquad \qquad r > R + \frac{\delta}{2} \nonumber \\
0 < h < \eta \,,\, 0 < K < 1 \qquad \qquad R - \frac{\delta}{2} \leq r \leq R + \frac{\delta}{2} 
\end{eqnarray} 
where $\delta$ is a length corresponding to the mass scale of the symmetry breaking.
As we shall see in section \ref{sec:collco}, describing the monopole in this way allows us to study the dynamics in terms of
just one collective coordinate $R$. The energy of the monopole then becomes
a simple polynomial in $R$. Furthermore, due to the spherical symmetry, $R$ is a function of time alone and so the
original field theoretic model in $3 + 1$ dimensions reduces to a one-dimensional problem involving $R(t)$.

We now proceed to elucidate the existence of monopole solutions which have the thin wall behavior described in 
the previous subsection. Redefining the couplings appearing in the potential (\ref{potential}) in terms of a mass
scale $\mu$ and expressing $\phi$ in terms of the profile function $h(r)$, we have
\begin{equation}
V = \frac{\tilde{\lambda}}{\mu^2}h^2\big(h^2 - \mu^2\tilde{a}^2\big)^2 + \tilde{\gamma}^2\mu^2h^2-\epsilon
\label{rescaledpotential}
\end{equation} 
where a tilde over a variable indicates that it is dimensionless. The vacuum expectation value of $\phi$ or $h$
then becomes $\tilde{\eta}\mu$, where
\begin{equation}
\tilde{\eta} = \sqrt{\frac{2\tilde{a}^2}{3} + \frac{\sqrt{\tilde{a}^4\tilde{\lambda}^2 - 
3\tilde{\gamma}^2\tilde{\lambda}}}{3\tilde{\lambda}}}.
\end{equation} 
The expression for $V$ can be rearranged as
\begin{equation}
V = \Big( \big(\tilde{\lambda}\tilde{a}^4 + \tilde{\gamma}^2\big)\mu^2 - 2\tilde{\lambda}\tilde{a}^2h^2 \Big)h^2
 + O(h^6).
\end{equation} 
The condition that $V$ is approximately quadratic in $h$ is given by
\begin{equation}
\frac{h^2}{\mu^2} << \frac{\tilde{\lambda}\tilde{a}^4 + \tilde{\gamma}^2}{2\tilde{\lambda}\tilde{a}^2}.\label{linear}
\end{equation} 
When the above condition is satisfied, $\partial V/ \partial h$ is linear in $h$. The equation of motion for
$h$ given in equation (\ref{eom}) can then be written as
\begin{equation}
h'' + \frac{2}{r}h' - \frac{2h}{r^2} - k^2h = 0
\label{msbe}
\end{equation} 
where $k^2 = (\tilde{\lambda}\tilde{a}^4 + \tilde{\gamma}^2)\mu^2$ and $K$ 
has been set to unity. 
Equation (\ref{msbe}) has the
form of the modified spherical Bessel equation whose general form is
\begin{equation}
z^2w'' + 2zw' - [z^2 + l(l+1)]w = 0
\label{msbegen}
\end{equation} 
for a function $w(z)$. The primes in the above equation denote derivatives with respect to $z$ and
equation (\ref{msbe}) is obtained from (\ref{msbegen}) with $l = 1$.

The solution of equation (\ref{msbe}) is
\begin{equation}
h(r) = C \Big( \frac{I_{3/2}(kr)}{\sqrt{kr}} \Big) = Ci_1(kr)
\end{equation} 
where $I_J$ is the modified Bessel function of the first kind of order $J$, $i_n$ is the modified spherical Bessel
function of the first kind of order $n$, and $C$ is
an arbitrary constant. The function $i_1(kr) \sim e^{kr}/(kr)$ for $kr\gg 1$ and is linear in $kr$ for small $kr\ll 1$. 
If we choose $C = e^{-k\xi }$ with arbitrarily large $k\xi $, we see that we can keep Eqn. (\ref{linear}) satisfied and hence stay with the linear equation for $h(r)$  for arbitrarily large $kr$.

\begin{figure}[!htp]
\begin{center}
\includegraphics[width=0.45\textwidth]{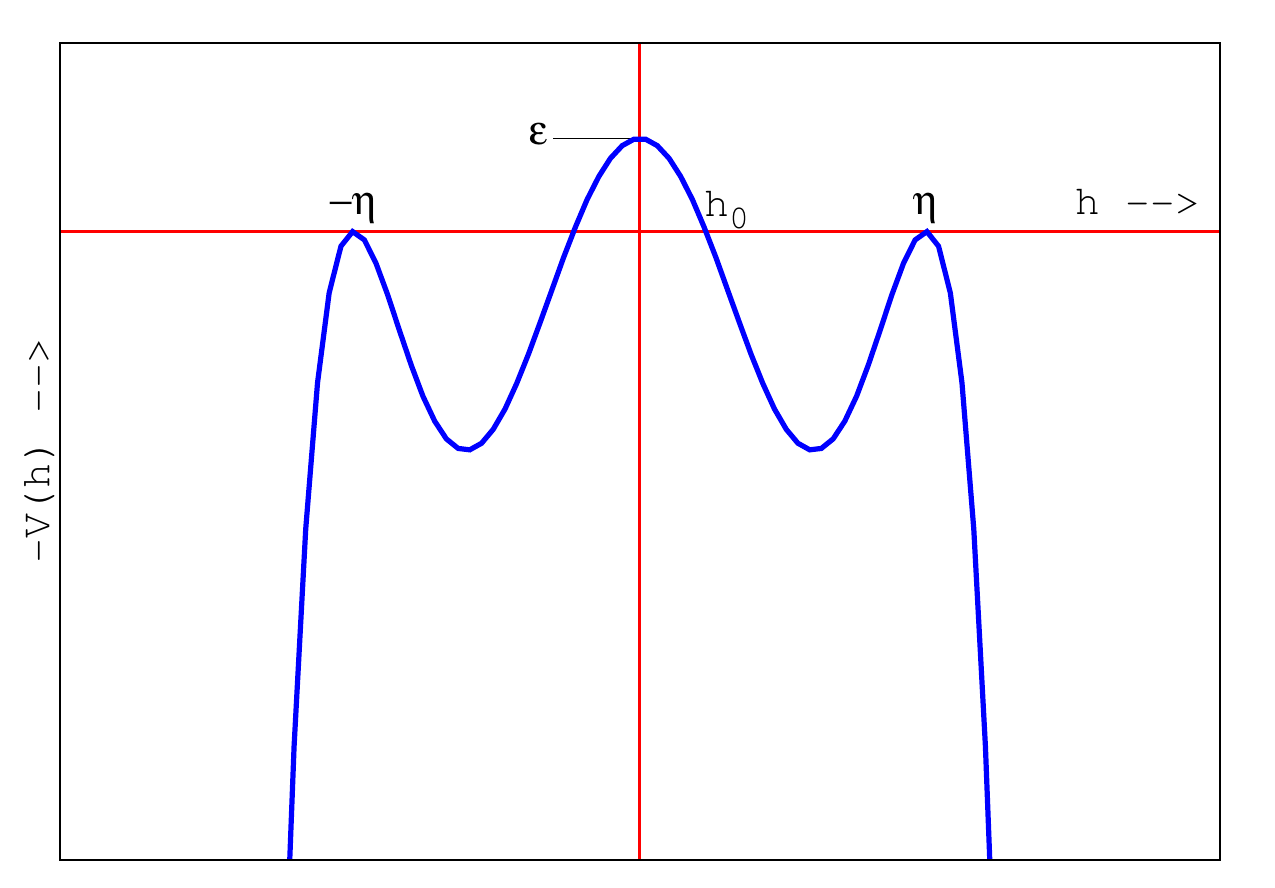}
\caption{The scalar potential $-V(h)$ which is the Euclidean space equivalent of the potential given
in (\ref{potential}). The potential has zeroes at $h = h_0$ and $h = \eta$.}
\label{fig4}
\end{center}
\end{figure}

The existence of the particular solution with $h(r) = \eta$ at $r=\infty$ can be proven using an argument similar to Coleman's, where he proved in a somewhat different context, the existence of a thin wall instanton, \cite{Coleman:1977py}.  We can reinterpret the equation for the monopole profile, Eqn. (\ref{monopole-equation}), as describing the motion of a particle whose position is denoted by $h(r)$ where $r$ is now interpreted as a time coordinate.  The particle moves in the presence of friction with a time dependent Stokes coefficient given by the second term in Eqn. (\ref{monopole-equation}) and a time dependent force given by the third term in Eqn. (\ref{monopole-equation}) (setting $K=1$), both of which are singular at $r=0$.   The particle also moves in the potential $-V(h)$, obtained by inverting the potential Eqn. (\ref{potential}),  as shown in figure \ref{fig4}.
The particle must start at $h = 0$ with a finite velocity  and must reach $h = \eta$ as $r\rightarrow\infty$.

We prove the existence of the solution that achieves $h = \eta$ at $r=\infty$ by
proving that initial conditions can be chosen so that the particle can undershoot or overshoot  $h = \eta$ for $r\rightarrow\infty$, depending on the choice of the initial velocity.  Then  by continuity there must exist an appropriate initial condition for which the particle exactly achieves $h=\eta$ at $r\rightarrow\infty$.  

In the following, we will assume that $K=1$ is always a good approximation.  Indeed, in Eqn, (\ref{monopole-equation}) the term dependent on $K$ is negligible for large $r$ no matter the value of $K$, while for small $r$, $K=1$ is a reasonable approximation.  On the other hand, Eqn. (\ref{eom}) for $K$, critically depends on the value of $h(r)\ne 0$, especially for large $kr$.   In that sense, the function $h(r)$ does not depend strongly on $K(r)$ whereas, $h(r)$ drives the behaviour of $K(r)$.

\subsection{Overshoot}
The existence of the overshoot can be proven by taking a sufficiently small value of $C$. As explained earlier, 
$C$ can be chosen small enough so that Eqn.(\ref{linear}) is valid even for large $kr$, hence the equation remains linear. If $kr$ is large enough, the friction 
term $(2/r) h'$ and the term $(2/r^2) h$ in the equation of motion can be neglected in any further evolution and the evolution can be thought of as conservative. Thus with such a choice
of $C$, $h$ increases to $\tilde h<h_0$ at a large value of $kr$ according to the linearised equation ($h_0$ is the zero crossing point of the potential, see Fig. (\ref{fig4})). The motion from then onwards is frictionless.  The particle has an energy $E > 0$ at $h = \tilde h$, thus it's energy is still positive when it reaches
$h = \eta$. As a result, it overshoots to $h > \eta$.

\subsubsection{Technical details}

\subsection{Undershoot}
To prove the existence of the undershoot, we start with the full equation for $h(r)$:
\begin{equation}
h'' + \frac{2}{r}h' - \frac{2}{r^2}h - \frac{\partial V}{\partial h} = 0
\end{equation}
which after multiplying both sides by $h'$ can be rewritten as
\begin{eqnarray}
\frac{d}{dr} \Big( \frac{1}{2}(h')^2 -V(h) \Big) &=& \label{energy}
-2h' \Big( \frac{h'}{r} - \frac{h}{r^2} \Big), \\
 &=& -2h'\big(\frac{h}{r}\big)'.\label{energy22}
\end{eqnarray} 
The quantity on the left hand side of Eqn. (\ref{energy}) can be thought of as the time derivative of the energy $E$.  In the linearised regime, it is easy to show that the right hand side is strictly negative for all $r$. It starts with a value of zero at $r=0$ and decreases essentially exponentially for large $kr$.  We can chose $C$, which amounts to choosing the initial velocity so that $h$ evolves according to the linearised equation until  $kr$ can be taken to be large.   However, in contrast to the case of the undershoot, we now require that $E$ becomes negative.  This means that the value of $C$ is taken larger than in the case of the overshoot.   $E$ is made up of two terms, the kinetic term which is positive semi-definite, and the potential term which becomes negative for $h>h_0$.  We impose conditions on the parameters so that $E$ becomes negative and consequently $h>h_0$ within the linearised regime.   Now if $kr$ is large enough, as before, the subsequent evolution will be conservative and since the total energy is negative, the subsequent evolution will never be able to overcome the hill at $h=\eta$ and the particle will undershoot. 
 
\subsection{Technical details}
To make the previous arguments more precise and rigorous, we note that when the condition Eqn. (\ref{linear})  is satisfied, the linear regime is valid and $V(h)$ is approximately quadratic in $h$, {\it ie.} $-V(h)\approx\epsilon - (1/2)k^2h^2$ and the equation of motion for $h$ is approximately
\begin{equation}
\frac{d}{dr} \Big( \frac{1}{2}(h')^2 + \epsilon - \frac{1}{2}k^2h^2 \Big)= -2h' \Big( \frac{h'}{r} - \frac{h}{r^2} \Big).
\end{equation} 
Using the properties of $i_1(kr)$ we can compute $E$ in the linear regime, we find for large $kr$
\beq
E\approx \epsilon - \frac{k^2C^2e^{2kr}}{4(kr)^3}\label{energy3}
\eeq
which can be evidently taken to be positive or negative by simply choosing the value of $C$.  Then in the subsequent evolution, where we can no longer rely on the linear evolution, the right hand side has two competing terms, the friction term, which only reduces the energy and the time dependent force term which tries to increase it.  The change in the energy for evolution between  $r_0$ and $r_f$ is given by the integral of the right hand side.
\subsubsection{Overshoot}
For the case of the overshoot, we use the expression Eqn. (\ref{energy22}) which gives
\beq
\Delta E=-2\int_{r_0}^{r_f} \, dr\,h'\big(\frac{h}{r}\big)'.
\eeq
Assuming that $h'(r)$ is positive, we will find an estimate for $h'(r)<v$.  Then
\bea
|\Delta E|&<&2v\left|\int_{r_0}^{r_f} \, dr\,\big(\frac{h}{r}\big)'\right|\\
&=&2v\left|\left(\frac{h(r_f)}{r_f}-\frac{h(r_0)}{r_0}\right)\right|\\
&<&2v\left|\left(\frac{\eta}{r_f}-\frac{h(r_0)}{r_0}\right)\right|
\eea
where we replaced $h(r_f)$ with $\eta$ since that is its largest possible value.  As long as $v$ is well behaved, as $r_0\to\infty$, $r_f>r_0$ thus the first term vanishes, while the second term can be made small by choosing the value of $C$ to be arbitrarily small.  Thus we see that $\Delta E\to 0$ and therefore the change in the energy is arbitrarily small.   Thus we necessarily obtain an overshoot since at $r=r_\eta$ such that $h(r_\eta)=\eta$, $V(\eta)=0$, hence the particle has a positive kinetic energy giving an overshoot.  

To get the value of $v$, we use Eqn. (\ref{energy})
\bea
\frac{d}{dr} \Big( \frac{1}{2}(h')^2 -V(h) \Big) &=& -2h' \Big( \frac{h'}{r} - \frac{h}{r^2} \Big), \\
&<&2\frac{hh'}{r^2}<\frac{(h^2)'}{r_0^2}.
\eea
Integrating both sides from $r_0$ to $r_f$ yields
\bea
(h'(r_f))^2&<&2\left( \frac{1}{r_0^2}\left( h^2(r_f)-h^2(r_0)\right)\right.\\
&+&\left. V(h(r_f))-V(h(r_0))+\frac{1}{2}(h'(r_0))^2\right).
\eea
Thus $v^2$ is given by
\bea
v^2&=&2\left( \frac{1}{r_0^2}\left( \eta^2-h^2(r_0)\right)\right.\\
&+&\left. \sup\left|V(h(r_f))-V(h(r_0))\right|+\frac{1}{2}(h'(r_0))^2\right)
\eea
which is a bounded function of $r_0$.
\subsubsection{Undershoot}
To prove the undershoot we use the expression Eqn. (\ref{energy}) which gives
\beq
\Delta E=-2\int_{r_0}^{r_f} \,dr\,\frac{{h'}^2}{r}+2\int_{r_0}^{r_f} \, dr\,\frac{{h'}h}{r^2}.
\eeq
Integrating the second term by parts we obtain
\bea
2\int_{r_0}^{r_f} \, dr\,\frac{{h'}h}{r^2}&=&\int_{r_0}^{r_f}  \, dr\,\left(\frac{h^2}{r^2}\right)'+\int_{r_0}^{r_f}  \, dr\,\left(\frac{2h^2}{r^3}\right)\\
&<& \left.\left(\frac{h^2}{r^2}\right)\right|_{r_0}^{r_f} -\eta^2\left.\left(\frac{1}{r^2}\right)\right|_{r_0}^{r_f}
\eea
where we obtain the inequality using the fact that we are only interested in the region $h\le \eta$.  

We now prove that this contribution to the energy cannot be sufficient push $h$ to $h>\eta$.  We take $r_0$ to be the value of $r$ as described after Eqn. (\ref{energy22}), where the energy becomes negative within the linearised regime with $kr_0\gg 1$.  We now assume there exists a value $r_f\equiv r_\eta$ for which $h(r_\eta)=\eta$.  Then
\bea
\Delta E &<&-2\int_{r_0}^{r_\eta} \,dr\,\frac{{h'}^2}{r} +\left.\left(\frac{h^2}{r^2}\right)\right|_{r_0}^{r_\eta} -\eta^2\left.\left(\frac{1}{r^2}\right)\right|_{r_0}^{r_\eta}\\
&<&\frac{\eta^2}{r_\eta^2}-\frac{h^2(r_0)}{r_0^2}-\eta^2\left(\frac{1}{r_\eta^2}-\frac{1}{r_0^2}\right)\\
&=&\frac{\eta^2-h^2(r_0)}{r_0^2}
\eea
which is an upper bound to the energy that can be added to the particle.  But now it is easy to see that this is insufficient for $kr_0$ large enough.  Indeed the energy of the particle at $r=r_0$ is obtained, via the linear regime, by Eqn. (\ref{energy3})
\beq
E\approx \epsilon - \frac{k^2C^2e^{2kr}}{4(kr)^3}\rightarrow\epsilon - 
k\frac{h^2(r_0)}{r_0} .
\eeq
This expression is negative.  Furthermore, if $kr_0$ is large enough, we will see that $\Delta E$ cannot provide enough energy to increase $E$ to zero, giving a contradiction to the existence of $r_\eta$.  To see this, we would require $|E|>\Delta E$ {\it ie.}
\beq
k\frac{h^2(r_0)}{r_0}-\epsilon >\frac{\eta^2-h^2(r_0)}{r_0^2}.
\eeq
The linear approximation assumes $h(r_0)\ll\eta$, hence we get
\beq
\frac{kh^2(r_0)}{r_0}-\frac{\eta^2}{r_0^2}>\epsilon
\eeq
reorganizing the terms, which for small enough $\epsilon$ simply implies
\beq
h^2(r_0)kr_0> \eta^2.
\eeq
Thus we get the the inequality sandwich
\beq
\frac{\eta^2}{kr_0}<h^2(r_0)<\eta^2.
\eeq
Using $h(r_0)\approx Ce^{kr_0}/2kr_0$ we can choose 
\beq
C= \frac{\eta 2kr_0}{e^{kr_0}r_0^{1/4}}
\eeq which gives
\beq
\frac{\eta^2}{kr_0}<\frac{\eta^2}{\sqrt{kr_0}}<\eta^2.
\eeq
It is obvious that for large enough $kr_0$ this is easily satisfied.  Thus we have established the existence of a choice of $C$ or initial velocity which contradicts the existence of $r_\eta$.

\section{Collective coordinate and the instantons}
\label{sec:collco}

\begin{figure}[!htp]
\begin{center}
\includegraphics[width=0.45\textwidth]{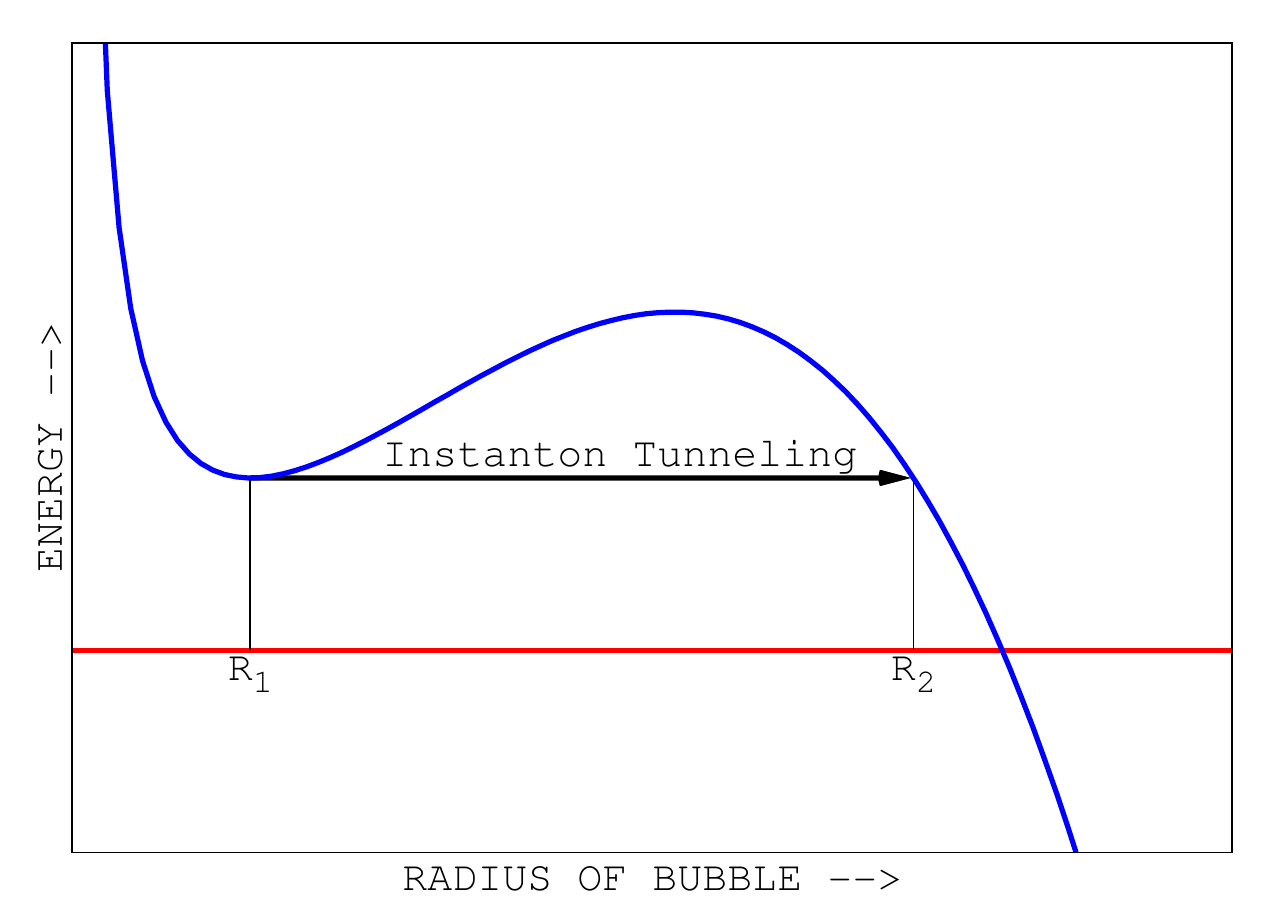}
\caption{The function $E(R)$ plotted versus bubble radius. The classically stable monopole solution has
$R = R_1$. This solution can tunnel quantum mechanically to a configuration with $R = R_2$ and then 
expand classically.}
\label{fig3}
\end{center}
\end{figure}

The potential $V(\phi)$ given in (\ref{potential}) can be normalized so that the energy density of the metastable
vacuum is vanishing whereas the energy density of the true vacuum is $-\epsilon$. By making use of the thin-wall
approximation, the expression for the total energy in the static case given in (\ref{staticenergy}) can be expressed as
\begin{eqnarray}
E = 4\pi &\Bigg[& \int_0^{R - \frac{\delta}{2}} dr \,r^2 V(h) 
+ \int_{R + \frac{\delta}{2}}^{\infty} dr \, \frac{1}{2e^2r^2} \nonumber \\ 
&+& \int_{R - \frac{\delta}{2}}^{R + \frac{\delta}{2}} dr \, \Big( \frac{(K')^2}{e^2} + \frac{(1-K^2)^2}{2e^2r^2} \nonumber \\
&+& \frac{1}{2}r^2(h')^2+ K^2h^2 + r^2V(h) \Big) \Bigg].
\label{energy2}
\end{eqnarray} 
In the above expression, we have made use of the fact that $V(h)$ is zero for $r > R + \frac{\delta}{2}$, $K = 1$  
for $r < R - \frac{\delta}{2}$, $K = 0$ for $r > R + \frac{\delta}{2}$, and that both the derivative terms and the 
term $K^2h^2$ are non-zero only when $R - \frac{\delta}{2} < r < R + \frac{\delta}{2}$. Since $\delta$ is small, the 
first integral on the right hand side of (\ref{energy2}) 
gives $-\alpha R^3$ where $\alpha = 4\pi\epsilon/3$ because $V(h) = -\epsilon$ in the domain of integration. The second
integral gives $C/R$ where $C = 2\pi/e^2$. The third integral is due to the energy of the wall and can be written as
$4\pi\sigma R^2$ where $\sigma$ is the surface energy density of the wall given by
\begin{eqnarray}
\sigma &=& \frac{1}{R^2} \int_{R - \frac{\delta}{2}}^{R + \frac{\delta}{2}} dr \, \Big( \frac{(K')^2}{e^2} +
\frac{(1-K^2)^2}{2e^2r^2} \nonumber \\
&+& \frac{1}{2}r^2(h')^2+ K^2h^2 + r^2V(h) \Big).
\label{wallenergy}
\end{eqnarray} 
We can thus write the total energy of the monopole as
\begin{equation}
E(R) = -\alpha R^3 + 4\pi\sigma R^2 + \frac{C}{R}.
\label{eofr}
\end{equation}
This function is plotted in figure \ref{fig3}. There is a minimum at $R = R_1$ and this corresponds to the classically stable 
monopole solution. This solution has a bubble of true vacuum in its core and the radius $R_1$ of this bubble is obtained
by solving $dE/dR = 0$. However, this monopole configuration can tunnel quantum mechanically through the finite barrier 
into a configuration with $R = R_2$ where $E(R_1) = E(R_2)$. Once this occurs, the monopole can continue to lose energy
through an expansion of the core since the barrier which was present at $R_1$ is no longer able to prevent this. 

We now proceed to determine the action of the instanton describing the tunneling from $R = R_1$ to $R = R_2$. 
In the thin wall approximation, the functions $h$ and $K$ can be written as
\begin{eqnarray}
h &=& h(r - R) \nonumber \\
K &=& K(r - R)
\label{thinwall}
\end{eqnarray} 
and the exact forms of the functions $h$ and $K$ will not be required in the ensuing analysis. The only requirement is that
both $h$ and $K$ change exponentially when their argument $(r - R)$ is small. An example of a function with this type of 
behaviour is the hyperbolic tangent function. The time derivative of $\phi$ can be written as
\begin{equation}
\dot{\phi}^a =  \hat{r}_a \frac{dh}{dR} \dot{R}.
\end{equation} 
From (\ref{thinwall}), since $(dh/dR)^2 = (dh/dr)^2$, we have
\begin{equation}
\frac{1}{2} \dot{\phi}^a \dot{\phi}^a = \frac{1}{2}  \left(\frac{dh}{dR}\right)^2 \dot{R}^2
= \frac{1}{2} \left(\frac{dh}{dr}\right)^2 \dot{R}^2.
\end{equation} 
Similarly, 
\begin{equation}
\dot{A}_{\mu}^a = \epsilon_{\mu ab} \hat{r}_b \left(\frac{-1}{er}\right) \frac{dK}{dR} \dot{R}
\end{equation} 
and 
\begin{equation}
\frac{1}{4} \dot{A}_{\mu}^a \dot{A}_{\mu}^a = \frac{1}{2e^2r^2} \left(\frac{dK}{dr}\right)^2 \dot{R}^2.
\end{equation} 
The Lagrangian can then be expressed as
\begin{equation}
L = 2\pi \int_0^{\infty} \Big( r^2 \left(\frac{dh}{dr}\right)^2\dot{R}^2 + 
\frac{1}{e^2}\left(\frac{dK}{dr}\right)^2\dot{R}^2 \Big) dr - E(R).
\label{totalLag}
\end{equation} 
From (\ref{eom}), for large r, the equation of motion of $h$ can be written as
\begin{equation}
h'' - \frac{\partial V(h)}{\partial h} = 0.
\end{equation} 
Multiplying both sides by $h'$ and integrating by parts with respect to $r$, one obtains
\begin{equation}
h' = \sqrt{2V(h)}.
\label{hprime}
\end{equation} 
Furthermore, since $dh/dr$ is non-vanishing only in the thin-wall, the value of $r$ in the first integral in 
(\ref{totalLag}) can be replaced by $R$ and we have
\begin{eqnarray}
\int_0^{\infty} dr \, r^2 \left(\frac{dh}{dr}\right)^2\dot{R}^2 &=&  R^2 \dot{R}^2 \int_0^{\infty} dr
\left(\frac{dh}{dr}\right)\sqrt{2V(h)} \nonumber \\
&=& R^2\dot{R}^2 S_1
\end{eqnarray} 
where
\begin{equation}
S_1 = \int_0^{\eta} dh\sqrt{2V(h)}.
\label{s1}
\end{equation} 
Defining 
\begin{equation}
S_2 = \frac{1}{e^2} \int_0^{\infty} dr \left(\frac{dK}{dr}\right)^2,
\end{equation} 
the Lagrangian (\ref{totalLag}) becomes
\begin{equation}
L = 2\pi\dot{R}^2( S_1 R^2  + S_2) - E(R) 
\end{equation} 
and the action can be written as
\begin{equation}
S = \int_{-\infty}^{\infty} dt \Big( 2\pi\dot{R}^2( S_1 R^2  + S_2) - E(R) \Big).
\end{equation}
In Euclidean space, the expression for the action becomes
\begin{equation}
S_E =  \int_{-\infty}^{\infty} d\tau \Big( 2\pi\dot{R}^2( S_1 R^2  + S_2) + E(R) \Big)
\label{euclideanaction}
\end{equation} 
where $\tau = it$ is the Euclidean time and $\dot{R}$ is the derivative with respect to $\tau$.
The instanton solution $R(\tau)$ which we are seeking
obeys the boundary conditions $R = R_1$ for $\tau = \pm \infty$, $R = R_2$ for $\tau = 0$, and
$dR/d\tau = 0$ for $\tau = 0$. It can be obtained by solving the equations of motion derived
from (\ref{euclideanaction}). However, the exact form for $R(\tau)$ will not be of interest here
since the decay rate of the monopole is determined ultimately from $S_E$ \cite{Coleman:1977py}.
The calculation of $S_E$ will be the subject of the next section.

\section{Bounce action}
\label{sec:decay}

In this section, we will derive an expression for bounce action $S_E$ for the monopole tunneling
and compare it with the bounce action for the tunneling of the false vacuum to the true vacuum
as discussed in \cite{Coleman:1977py} with no monopoles present.
From (\ref{euclideanaction}), the equation of motion for $R$ can be written
\begin{equation}
(R^2S_1 + S_2)\ddot{R} + S_1R\dot{R}^2 - \frac{1}{4\pi}\frac{\partial E}{\partial R} = 0.
\end{equation} 
Multiplying both sides by $\dot{R}$, the equation of motion assumes the form
\begin{equation}
\frac{d}{dt} \left[ \frac{1}{2}(S_2 + R^2S_1)\dot{R}^2 - \frac{E(R)}{4\pi} \right] = 0.
\end{equation} 
The term in the square brackets is a constant of motion and can be taken to be zero with loss of
generality. Setting this constant to zero gives
\begin{equation}
E(R) = 2\pi (S_2 + S_1R^2)\dot{R}^2.
\label{com}
\end{equation} 
Substituting this in (\ref{euclideanaction}), we have
\begin{equation}
S_E = \int_{-\infty}^{\infty} d\tau \, 4\pi(S_2 + S_1R^2)\dot{R}^2.
\end{equation} 
Solving for $\dot{R}$ from (\ref{com}) and using this in the above equation yields
\begin{eqnarray}
S_E &=& \int_{-\infty}^{\infty} d\tau (\frac{dR}{d\tau}) 4\pi (S_2 + S_1R^2) \dot{R} \nonumber \\
&=& \sqrt{32\pi} \int_{R_1}^{R_2} dR \sqrt{(S_2 + S_1R^2)E(R)}.
\end{eqnarray} 
Using the expression for $E(R)$ given in (\ref{eofr}) and neglecting $S_2$ in comparison to
$S_1R^2$, the euclidean action of the bounce solution can be written
\begin{equation}
S_E =  A \int_{R_1}^{R_2} dR \sqrt{-(\alpha R^5 - 4\pi\sigma R^4 - CR + E_0R^2) }
\end{equation}
where $A = \sqrt{32\pi S_1}$.
In deriving the above expression, the constant $E_0 = E(R_1)$ was subtracted from the expression for
$E(R)$ in (\ref{eofr}) so that the bounce has a finite action. Pulling out a factor of $R$ from the
square root in the integrand, we have
\begin{equation}
S_E =  A \int_{R_1}^{R_2} dR \sqrt{R} \sqrt{-J} 
\end{equation}
where $J = \alpha R^4 - 4\pi\sigma R^3 - C + E_0R$. The function $J$ has a double root at $R = R_1$,
a positive root at $R = R_2$, and a negative root at $R = R_3$. Since we are working with $\epsilon$ 
small and $\alpha = 4\pi\epsilon/3$, we can neglect the term containing $\alpha$ while solving $dE/dR = 0$
and obtain $R_1 \approx (4e^2\sigma)^{-1/3}$.  To find $R_3$ we also neglect the term containing $\alpha$, and substituting for $E_0$ in terms of the solution for $R_1$, we get a cubic equation for $R_3$, which can be exactly factored, giving $R_3=-2R_1$.  Finally, to solve for $R_2$, we solve $J = 0$ neglecting the 
constant and linear term in $R$ since $R_2$ is large, obtaining $R_2 \approx 4\pi\sigma/\alpha=3\sigma/\epsilon$. 

Factoring $J$, we have
\begin{eqnarray}
S_E &=&  A \sqrt{\alpha} \int_{R_1}^{R_2} dR \sqrt{R} \sqrt{-(R-R_1)^2(R-R_2)(R-R_3)} \nonumber  \\  
 &=&  A \sqrt{\alpha} \int_{R_1}^{R_2} dR \sqrt{R}(R-R_1) \sqrt{-(R-R_2)(R-R_3)} \nonumber \\ 
 &=&  A \sqrt{\alpha} R_2^{7/2} \frac{2}{105} \Big(1 - \frac{R_1}{R_2}\Big)^{5/2} \, I\Big(\frac{R_1}{R_2},\frac{R_3}{R_2}\Big)\label{59}\\
 &=&\sqrt{32\pi S_1}\sqrt{4\pi\epsilon/3}R_2^{7/2} \frac{2}{105} \Big(1 - \frac{R_1}{R_2}\Big)^{5/2} \, I\Big(\frac{R_1}{R_2},\frac{R_3}{R_2}\Big).\nonumber
\end{eqnarray} 
Here $I$ is a dimensionless function of $R_1/R_2$ and $R_3/R_2$ which is finite everywhere in the domain 
$[R_1,R_2]$ and is obtained from the integral defined in Eqn, (\ref{59}) removing the factor of $(1-(R_1/R_2))^{(5/2)}$ and $R_2^{7/2}$ and some numerical factors.  It is expressible in terms of elliptic integrals and its explicit expression is not illuminating.  As $S_1$ has dimensions of $\mu^3$ and $\epsilon$ has dimensions of $\mu^4$, the expression is dimensionless, as expected.  Substituting the value of $R_2$ in $S_E$,
\begin{equation}
S_E = \frac{144\,\pi}{35} \sqrt{2\,S_1} \frac{\sigma^{\frac{7}{2}}}{\epsilon^3} \Big(1 - \frac{R_1}{R_2}\Big)^{5/2} 
\, I\Big(\frac{R_1}{R_2},\frac{R_3}{R_2}\Big).
\label{bounce1}
\end{equation} 
For small $\epsilon$, the term containing $\tilde{\gamma}$ in the potential (\ref{rescaledpotential}) can be 
neglected. Using equation (\ref{s1}) and the fact that $\eta = \tilde{a}\mu$ when $\tilde{\gamma} = 0$,
\begin{eqnarray}
S_1 &=& \frac{\sqrt{2\tilde{\lambda}}}{\mu} \int_0^{\tilde{a}\mu} dh \,\big( h(h^2 - \mu^2\tilde{a}^2) \big) \\
 &=& \sqrt{\frac{\tilde{\lambda}}{8}} \tilde{a}^4 \mu^3.
\label{s1value}
\end{eqnarray} 
The value of $\sigma$ can be obtained from equation (\ref{wallenergy}) by noting that the terms multiplying $r^2$ 
are large compared to the terms independent of $r$ and the term multiplying $1/r^2$. Since $\delta$ is small, we
can write $r = R$ and equation (\ref{wallenergy}) becomes
\begin{equation}
\sigma = \int_{R - \frac{\delta}{2}}^{R + \frac{\delta}{2}} dr \, \Big(\frac{1}{2}(h')^2 + V(h) \Big).
\end{equation} 
Substituting for $h'$ from equation (\ref{hprime}), $\sigma$ becomes
\begin{eqnarray}
\sigma &=& \int_{R - \frac{\delta}{2}}^{R + \frac{\delta}{2}} dr \, (h')^2 \\ 
 &=&  \int_{0}^{\eta} dh \, (h') \\ 
 &=&  \int_{0}^{\eta} dh \, \sqrt{2V(h)} \\ 
 &=&  S_1.
\label{sigma}
\end{eqnarray}
Using (\ref{sigma}) and (\ref{s1value}) in (\ref{bounce1}) yields
\begin{eqnarray}
S_E &=& \frac{144\,\pi\sqrt{2}}{35} \frac{S_1^4}{\epsilon^3} \Big(1 - \frac{R_1}{R_2}\Big)^{5/2} \, I\Big(\frac{R_1}{R_2},\frac{R_3}{R_2}\Big) \\ 
 &=& \frac{9\sqrt{2}\,\pi}{140} \tilde{\lambda}^2\tilde{a}^{16} \frac{\mu^{12}}{\epsilon^3} \Big(1 - \frac{R_1}{R_2}\Big)^{5/2} 
\, I\Big(\frac{R_1}{R_2},\frac{R_3}{R_2}\Big)
\label{finalbounce}
\end{eqnarray}
as the final value of the bounce action. From the values of $R_1$ and $R_2$, we have
\begin{eqnarray}
\frac{R_1}{R_2} &=& \frac{1}{e^{2/3}} \frac{1}{(4\sigma)^{1/3}} \frac{\epsilon}{3\sigma} \\
 &=&  \frac{1}{(\tilde{\lambda}e)^{2/3}} \big( \frac{16}{27} \big)^{1/3} \frac{\epsilon}{\tilde{a}^{16/3} \mu^4}
\label{r1overr2}
\end{eqnarray} 
where the value of $\sigma$ has been expressed in terms of the couplings appearing in the potential using
equations (\ref{sigma}) and (\ref{s1value}). From the expression given in (\ref{finalbounce}), it is evident 
that the bounce action $S_E$ is zero when $R_1 = R_2$ as expected. With $\epsilon$ small, $R_1/R_2$ is small, 
but it is interesting to note that variations in the couplings can reduce the bounce action. For example, a 
reduction in the $U(1)$ gauge coupling $e$ has the effect of increasing the monopole mass and of reducing the bounce action.

We now compare our answer with the well known formula of \cite{Coleman:1977py} relevant to 
homogeneous nucleation, i.e. tunneling 
of the translation invariant false vacuum to the true  vacuum. Denoting this bounce to be $B_0$,
\begin{eqnarray}
B_0 &=& \frac{27\pi^2}{2} \frac{S_1^4}{\epsilon^3} \\
 &=&  \frac{27\pi^2}{128} \tilde{\lambda}^2 \tilde{a}^{16} \frac{\mu^{12}}{\epsilon^3}.
\end{eqnarray} 
Comparing this expression with our bounce $B \equiv S_E$ for the monopole assisted tunneling 
given in (\ref{finalbounce}), we see that 
\begin{equation}
B = \frac{32\sqrt{2}}{105 \, \pi} B_0 \, \Big(1 - \frac{R_1}{R_2}\Big)^{5/2} \, I\Big(\frac{R_1}{R_2},\frac{R_3}{R_2}\Big).
\end{equation} 
We see that unlike the homogeneous case, the bounce can parametrically become indefinitely
small and vanish in the limit $ R_1 \rightarrow R_2 $. The interpretation of this limit
is that the very presence of a monopole in this parameter regime implies the unviability
of a state asymptotically approaching the vacuum deduced by a naive use of the effective potantial. 
If the parameters in the effective potential explicitly depend on external variables
such as temperature, it may happen that the limit $ R_1 \rightarrow R_2 $ is reached at a 
critical value of this external parameter. In this case, as the external parameter gets tuned to 
this critical value, the monopoles will become sites where the true vacuum is nucleated without 
any delay and the indefinite growth
of such bubbles will eventually convert the entire system to the true vacuum without
the need for quantum tunneling. Such a phenomenon may be referred to as a \textit{roll-over
transition} \cite{Yajnik:1986wq} characterised by the relevant critical value.

\section{Monopole decay in a supersymmetric SU(5) GUT model}
\label{sec:gutmodel}
The results of this work have direct relevance to a supersymmetric $SU(5)$ model studied in \cite{Bajc:2008vk}
in which supersymmetry symmetry breaking is sought directly through O'Raifeartaigh type breaking. The Higgs sector, which contains two adjoint scalar superfields $\Sigma_1$ and $\Sigma_2$ and the 
superpotential, including leading non-renormalizable terms, is of the form
\begin{eqnarray}
W &=& Tr\left[\Sigma_{2}\left(\mu \Sigma_{1} + \lambda \Sigma_{1}^{2} +
\frac{\alpha_{1}}{M}\Sigma_{1}^{3} +
\frac{\alpha_{2}}{M}Tr(\Sigma_{1}^2)\Sigma_{1} \right) \right] \nonumber \\
&=& \sigma_{1}\sigma_{2} \, \left(\mu - \frac{\lambda}{\sqrt{30}}\sigma_{1} 
+ (7\alpha_{1} + 30\alpha_{2})\frac{\sigma_{1}^{2}}{30M}\right)
\label{superpot}
\end{eqnarray}
where $\sigma_1$ and $\sigma_2$ are selected components of $\Sigma_1$ and $\Sigma_2$
respectively, relevant to the symmetry breaking. Two mass scales appear in the superpotential,
$\mu$ and $M$, the latter being a larger mass scale whose inverse powers determine
the magnitudes of the coefficients of the non-renormalizable terms. The scalar potential
derived from this superpotential can be written as
\begin{eqnarray}
V &=& \Big( \mu \sigma_{1} - \frac{\lambda \sigma_{1}^2}{\sqrt{30}} +
\frac{7\alpha_{1} \sigma_{1}^3}{30\,M} + \frac{\alpha_{2}\sigma_{1}^3}{M}
\Big)^2 \nonumber  \\
&+& \Big( \sigma_2\Big(\mu - \frac{2\lambda\sigma_{1}}{\sqrt{30}}
+ \frac{(7\alpha_{1} + 30\alpha_{2})}{10\,M}\sigma_{1}^2 \Big)\Big)^2.
\label{scalpot}
\end{eqnarray}
In  \cite{Kumar:2009pr}, monopole solutions were shown 
to exist in this model  and the classical instability of the vacuum structure of this theory in the 
presence of such monopoles was discussed. 

Thin walled monopoles can be obtained in this model under the condition
\beq
\frac{\sigma_1}{\mu}\ll \frac{\sqrt{30}}{2\lambda}
\eeq
which is equivalent to the condition in Eqn. (\ref{linear}), and hence the results of this paper could be applied directly there.  In \cite{Kumar:2009pr} the region of parameter space studied did not coincide with this condition, and thus the monopoles were not thin walled.  The monopoles were classically unstable when $\epsilon\sim M^4$ was increased beyond a critical value.  We can recover this behaviour from Eqn. (\ref{finalbounce}) as $\epsilon$ is increased, however it is important to note that our approximation in this paper becomes invalid for large enough $\epsilon$.

\section{ Discussions and conclusions }
\label{sec:conc}
We have calculated the decay rate for so-called false monopoles in a simple model with a hierarchical structure of symmetry breaking.  The toy model that we use has a breaking of $SU(2)$ to $U(1)$ which is the false vacuum, which in principle happens at a higher energy scale, and then a true vacuum which has no symmetry breaking.  The symmetry broken false vacuum admits magnetic monopoles.  The false vacuum can decay via the usual creation of true vacuum bubbles \cite{Coleman:1977py}, however we find that this decay can be dramatically enhanced in the presence of magnetic monopoles.  Even though the false vacuum is classically stable, the magnetic monopoles can be unstable.  At the point of instability, the monopoles are said to dissociate.  This corresponds to an evolution where the core of the monopole, which contains the true vacuum, dilates indefinitely, \cite{Steinhardt:1981ec,Hosotani:1982ii,Steinhardt:1981mm}.  However, before the monopoles become classically unstable, they can be rendered unstable from quantum tunneling.  We have computed the corresponding  rate  and find that as we approach the regime of classical instability, the exponential suppression vanishes.  The tunneling amplitude behaves as

\beq
\frac{\Gamma}{V}\sim \left(\frac{\kappa}{2}\right) \exp\left\{\frac{16}{105}\sqrt{\frac{2 S_1\pi^2\epsilon}{3}}
\mathcal{F}(R_1, R_2, R_3)\right\} 
\eeq
with
\beq
\mathcal{F}(R_1, R_2, R_3)= R_2^{7/2} \Big(1 - \frac{R_1}{R_2}\Big)^{5/2} \, I\Big(\frac{R_1}{R_2},\frac{R_3}{R_2}\Big)
\eeq
where $\kappa$ contains the determinantal and zero mode factors, and $I$ is defined in Eqn. (\ref{59}).  In the limit that $R_1\to R_2$ the tunneling rate is unsuppressed while the homogeneous tunneling rate for the nucleation of true vacuum bubbles as found by Coleman \cite{Coleman:1977py} still remains suppressed.  Hence in this limit, the classical false vacuum is classically stable, but subject to quantum instability through the nucleation of true vacuum bubbles, but the rate for such a decay can be quite small.  However the existence of magnetic monopole defects render the false vacuum unstable, and in the limit of large monopole mass, the decay rate is unsuppressed.  

\maketitle

\section{ACKNOWLEDGEMENTS}

We thank NSERC, Canada for financial support. The visit of BK was made possible by a grant from
CBIE, Canada. The research of UAY is partly supported by a grant from DST, India. The authors
would like to thank R. MacKenzie and P. Ramadevi for useful comments regarding this work.


\bibliographystyle{apsrev}
\bibliography{ref}

\end{document}